\documentclass[preprint]{revtex4-1}

\usepackage[usenames,dvipsnames]{color}
\usepackage{amssymb}
\usepackage{amsmath}
\usepackage{hyperref}
\usepackage{graphicx}
\usepackage{subfigure}
\usepackage{xspace}

\usepackage{lipsum}

\begin{document}

\title{Self-assembly of finite-sized colloidal aggregates}

\author{Pritam Kumar Jana$^{\ast}$\textit{$^{a,b}$} and Bortolo Matteo Mognetti$^{\ddag}$\textit{$^{a}$}}

\address{
$^{a}$Universit\'e Libre de Bruxelles (ULB), Interdisciplinary Center for Nonlinear Phenomena and Complex Systems, Campus Plaine, CP 231, Blvd.\ du Triomphe, B-1050 Brussels. \\
$^{b}$~Georg--August Universit\"{a}t G\"{o}ttingen, Institute for Theoretical Physics, 37077 G\"{o}ttingen, Germany.\\
$^{\ast}$pritamkumar.jana@theorie.physik.uni-goettingen.de $^{\ddag}$bmognett@ulb.ac.be
}

\begin{abstract}
One of the challenges of self-assembling finite-sized colloidal aggregates with a sought morphology is the necessity of precisely sorting the position of the colloids at the microscopic scale to avoid the formation of off-target structures. Microfluidic platforms address this problem by loading into single droplets the exact amount of colloids entering the targeted aggregate. Using theory and simulations, in this paper, we validate a more versatile design allowing us to fabricate different types of finite-sized aggregates, including colloidal molecules or core-shell clusters, starting from finite density suspensions of isotropic colloids in bulk. In our model, interactions between particles are mediated by DNA linkers with mobile tethering points, as found in experiments using DNA oligomers tagged with hydrophobic complexes immersed into supported bilayers. By fine-tuning the strength and the number of the different types of linkers, we prove the possibility of controlling the morphology of the aggregates, in particular, the valency of the molecules and the size of the core-shell clusters. In general, our design shows how multivalent interactions can lead to microphase separation in equilibrium conditions. 
\end{abstract} 

\maketitle

\section{Introduction}

Designing new bottom-up fabrication procedures leading to the formation of finite colloidal clusters with tailored morphology is of great importance in many Materials Science applications, including energy conversion, catalysis, and sensing (e.g.~\cite{Kamat2007,Fang2009}).

The use of functionalized colloids featuring selective interactions, in which pairs of colloids only interact when carrying complementary moieties, enables controlling the morphological properties of both crystalline and disordered aggregates at the single-particle level \cite{GerthChemComm2017}. In particular, DNA oligomers have been broadly used to mediate interactions between colloids\cite{MirkinNature1996,AlivisatosNature1996} because of the possibility of engineering high dimensional \cite{LowensohnPRX2019} and state-dependent interaction matrices \cite{JonesScience2015,RogersNatRevMat2016,Knorowski2011materials}. Such versatility arises from the selectivity of the Watson-Crick pairing as well as from precise control of the thermodynamics and kinetics of the hybridization of complementary single-stranded DNA oligomers \cite{A-UbertiNatMat2015,RogersScience2015}.

So far, uniformly coated colloids have been used mainly to self-assemble extended structures like crystals \cite{NykypanchukNatMat2008,AuyeungNature2014,Ducrot_NatMat_2017,Wang_NatComm_2017,Liu_Science_2016} or gels \cite{VarratoPNAS2012,DiMicheleSoftMatter2014}. Finite aggregates, like colloidal molecules \cite{VanBlaaderenScience2003,ManoharanScience2003}, have been fabricated using directional interactions as in systems of patchy particles  \cite{WangNature2012,SciortinoPCCP2010,PreislerJPCB2013,halverson2013DNA,MorphewACSNano2018}. Patchy particles are expensive building blocks that are often synthesized starting from pre-assembled clusters of isotropic colloids \cite{Pawar2010,MerindolChemistry2019}. Clusters of isotropic colloids\cite{Zeravcic_2014_PNAS,Arkus_2009_PRL,Meng_2010_Science,ZeravcicRMP2017} are usually fabricated in controlled environments, for instance, using microfluidic platforms in which drops are loaded with the exact amount of particles entering the targeted structure \cite{LeeLabChip2014}. However, the necessity of using direct interventions to sort colloids at the microscopic scale challenges the large scale production of finite-sized aggregates. In this work, we address this problem and present a single pot, bottom-up scheme to self-assemble different types of colloidal clusters with tailored size starting from colloidal suspensions at finite density.

\begin{figure*}[h!]
\centering
 \includegraphics[scale=0.35]{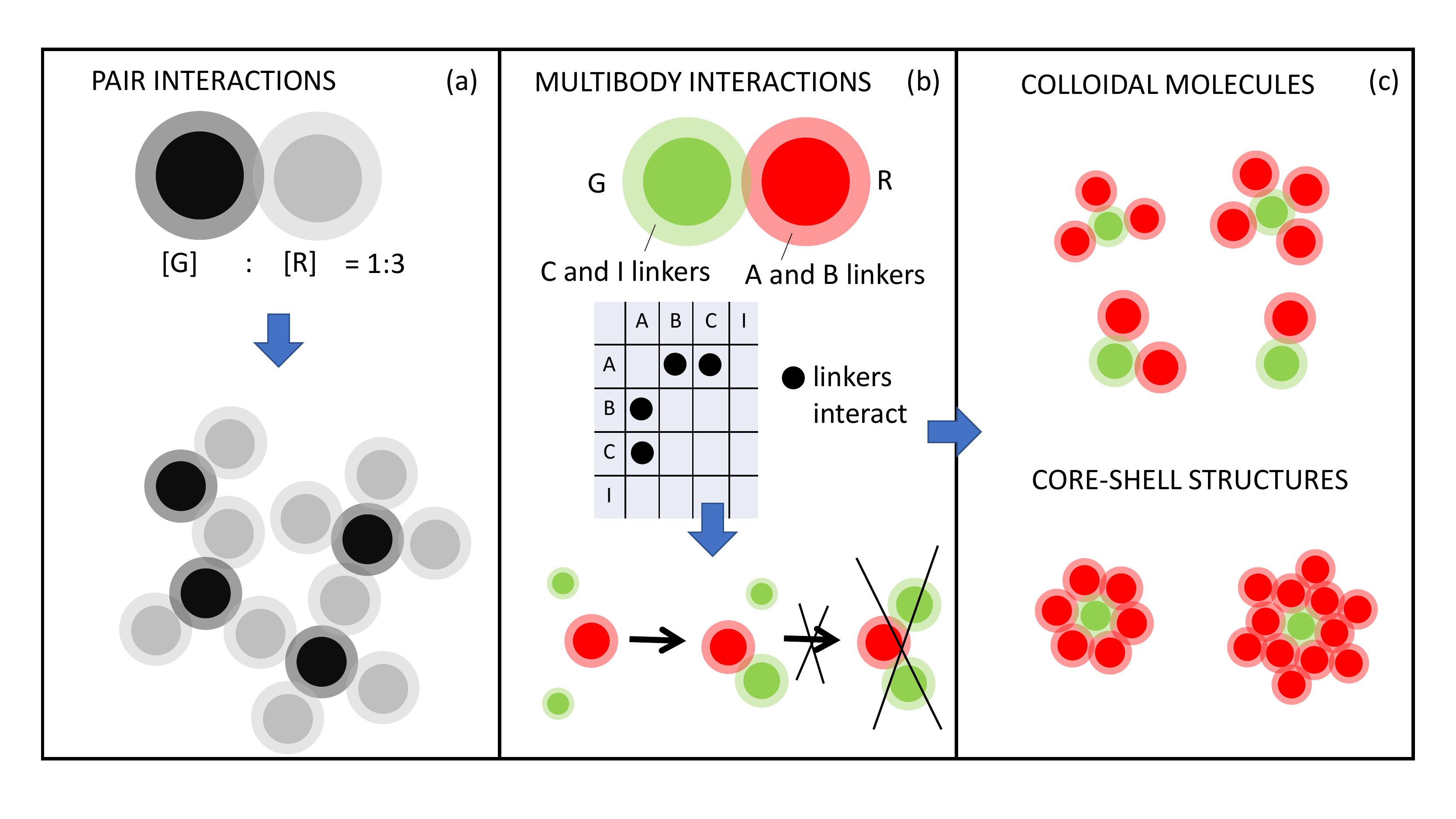} 
\caption{
{\bf Designing self--assembly of finite--sized aggregates.} (a) Suspensions of particles interacting through pair-interactions will generally aggregate into extended structures. (b) Top: We consider a binary system in which G (R) particles carry linkers of type C and I  (A and B). B and C linkers compete to bind A linkers (as marked in the middle table) while inert linkers do not react. Bottom: The effective interaction between particles is multibody, allowing to destabilize configurations in which R particles cross-link two G particles. (c) Using the building blocks defined in panel (a), in this paper we yield colloidal molecules with controllable valency (top) and core-shell structures (bottom).  
}\label{FigIntro}
\end{figure*}

We study suspensions made of two types of colloids (in the following tagged with G and R). Aiming at fabricating colloidal molecules (for instance GR$_3$), it would be tempting to engineer an attractive pair-interaction between G and R and prepare suspensions with a stoichiometric ratio equal to [G]:[R]=1:3 (where [G] and [R] are the concentrations of G and R colloids, respectively). However, this setting would unlikely lead to the sought type of colloidal molecule because local density fluctuations would result in a broad distribution of the number of  particles {\em per} cluster. Moreover, different {\em R} particles would likely crosslink different {\em G} particles, and the system would fail even to self-assemble finite-sized aggregates (see Fig.~\ref{FigIntro}a bottom).

To assemble finite clusters, here we engineer multibody interactions \cite{MognettiRPP2019} to design a system in which configurations  with an R particle crosslinking two G particles are thermodynamically unstable (see Fig.~\ref{FigIntro}b bottom). We consider particles functionalized by DNA linkers free to move on the surface of the colloids. Our model mimics systems of colloid supported bilayers \cite{Boxer_JACS_2005,Beales_BJ_2009,TroutierAdvCollIntSci20071,VdMeulenJACS2013,Kraft_Nanoscale_2017,RinaldinSoftMatter2018} or emulsions \cite{Jasna_PNAS_2012,Hadorn_PNAS_2012,Jasna_SoftMatter_2013} functionalized by DNA strands tethered to the liquid interface through a hydrophobic molecule. Particles with mobile linkers feature peculiar responsiveness and have been used, for instance, to control the aggregates’ valency \cite{A-UbertiPRL2014} or to engineer the self-assembly pathway through the use of cascade reactions \cite{HalversonJCP2016,ParoliniACSNano2016,DiMicheleJCP2016,zhang2017sequential,zhang2018multivalent,JanaNanoscale2019}.  We validate our model using state-of-the-art multivalent theories \cite{VarillyJCP2012,A-UbertiJCP2013,DiMicheleJCP2016,MognettiRPP2019,TitoJCP2016,CurkJPCM2020,A-UbertiFronNano2019,melting-theory1,Martinez-Veracoechea_PNAS_2011,XuShawBJour2016,KitovJACS2003} and simulations \cite{A-UbertiPRL2014,PetitzonSoftMatter2016,JanaPRE2019,A-UbertiFronNano2019} which have been corroborated in previous investigations \cite{JanaNanoscale2019}. Our design relies on competition between selective attractive forces and repulsive steric interactions engendered by inert strands (see Fig.~\ref{FigIntro}b top), as used recently by Angioletti-Uberti et {\em al.} to control the valency of percolating aggregates \cite{A-UbertiPRL2014}. In particular, by balancing the strength of the moieties tethered to G and R colloids with the number of inert linkers grafted to G particles, we achieve self-assembly of colloidal molecules with a controllable valency (see Fig.~\ref{FigIntro}c top). Moreover, our model also allows for exquisite control over the interactions between R particles. As a result, we design systems that self--assemble core--shell clusters in which one G molecule is wrapped by a controllable number (in this work two) of shells of R particles (see Fig.~\ref{FigIntro}c down). This goal is achieved using a strategy similar to what we recently used to self-assemble crystals with finite thickness \cite{JanaNanoscale2019}. In the present study, the size of the core-shell aggregates is limited by the fact that R particles interact more weakly when found in the outer shells of the clusters so that beyond a certain size the aggregate becomes unstable.

\section{Model and Methods}\label{SecModel}

\begin{figure}[h]
\centering
 \includegraphics[scale=0.25]{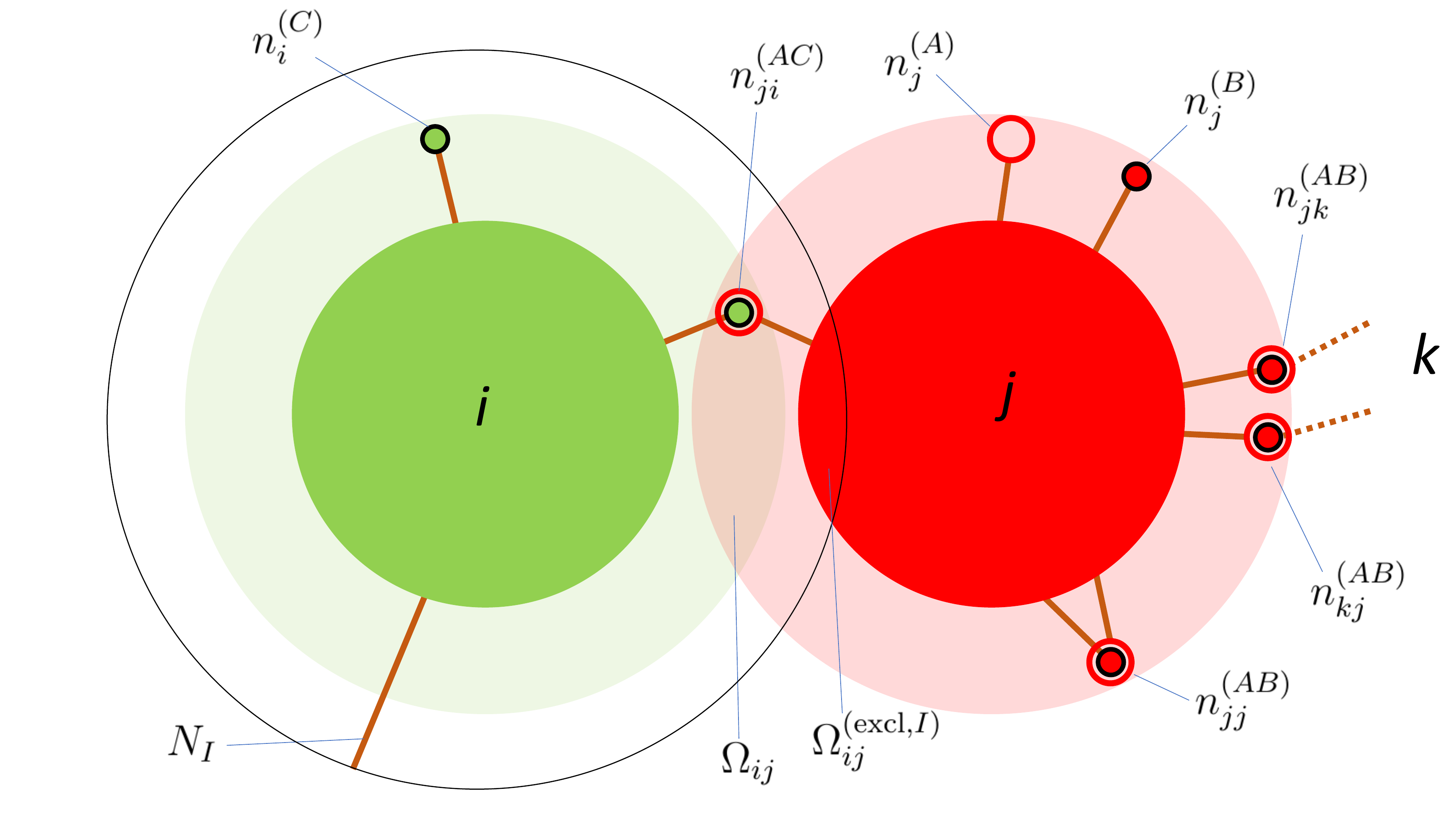} 
\caption{
{\bf Model parameters.} $N_I$ and $n^{(X)}_i$ are, respectively, the number of inert linkers tethered to each G particle and the number of unpaired linkers of type X (X=A, B, or C) tethered to particle $i$. $n^{(XY)}_{ij}$ is the number of linkages between linkers X and Y tethered, respectively, to particle $i$ and $j$. The number of linkages, e.g.~$n^{(AC)}_{ji}$, is controlled by the volume available to pairs of reacted tips ($\Omega_{ij}$, see Eq.~\ref{Eq:linkages}). $\Omega_{ij}$ directly affects the attraction force between particles (see the first term of Eq.~\ref{Eq:forces}). $\Omega^{(\mathrm{excl},I)}_{ij}$ is the volume excluded to each inert linker tethered to particle $i$ by the presence of colloid $j$. Similarly, $\Omega^{(\mathrm{excl})}_{ij}$ is the volume excluded to linkers A, B and C by $j$ ($\Omega^{(\mathrm{excl})}_{ij}=0$ in the figure). $\Omega^{(\mathrm{excl})}$ and $\Omega^{(\mathrm{excl},I)}$ control the osmotic terms resulting in particle-particle repulsions (see 2$^\mathrm{nd}$, 3$^\mathrm{rd}$, and 4$^\mathrm{th}$ term in Eq.~\ref{Eq:forces}). 
}\label{FigModel}
\end{figure}
We consider a binary system with two types of particles (R and G, see Fig.~\ref{FigModel}). Particles carry rigid spacers (as obtained using few tens of nanometer long segments of double-stranded DNA) tipped by reactive sequences. As anticipated by Fig.~\ref{FigIntro}b, R particles carry sequences of type A and B, while G particles are functionalized by a third reactive sequence (C) and inert linkers (I). Linkers of type B and C compete to bind A, while all other pairings are forbidden (see the table in Fig.~\ref{FigIntro}b). 
\\
Interactions between particles are estimated using a free energy $f$ accounting for all possible microstates of the linkers at a given colloids' configuration, $\{ {\bf r}_i \}_{i=1,\cdots,N_p}$, where $N_p$ is the total number of particles \cite{VarillyJCP2012,A-UbertiPRL2014,DiMicheleJCP2016,MognettiRPP2019}.
The free energy is decomposed into a multivalent part $f_\mathrm{multi}$ summing all contributions due to the reactive linkers, and a term due to inert linkers $f_\mathrm{inert}$, $f=f_\mathrm{multi}+f_\mathrm{inert}$. We calculate $f_\mathrm{multi}$ using portable expressions \cite{A-UbertiPRL2014,DiMicheleJCP2016} depending on the number of bridges and loops made by pairs of bound linkers (in the following, we generally label pairs of reacted linkers with linkages). 
We define by $n^{(XY)}_{ij}$ the number of linkages made by linkers of type $X$ tethered to particle $i$ bound to linkers of type $Y$ tethered to particles $j$  (see Fig.~\ref{FigModel}), where $X,\,Y=A,\, B,$ or $C$. 
In Fig.~\ref{FigModel}, $n^{(X)}_i$ counts the number of free linkers of type $X$ on particle $i$.
Each R particle carries $N_R$ linkers of type $A$ and $N_R$ of type $B$, while each $G$ particle $N_C$ linkers of type $C$ and $N_I$ inert linkers. 
Defining $\sigma_i=0$/$\sigma_i=1$ if particle $i$ is of type R/G, we obtain the following compact expression of the multivalent free--energy \cite{A-UbertiPRL2014}:
\begin{eqnarray}
{f_\mathrm{multi} \over k_B T} &=& \sum_i \Bigg[\delta_{\sigma_i,0} N_R \log {n^{(A)}_i n^{(B)}_i \over (N_R)^2} +\delta_{\sigma_i,0} n^{(AB)}_{ii} + \delta_{\sigma_i,1} N_C \log {n^{(C)}_i \over N_C} \Bigg]  
\nonumber \\
&& +\sum_{i<j} n_{ij}+{f^{A,B,C}_\mathrm{rep} \over k_B T}
\label{Eq:fmulti}
\end{eqnarray}
where $n_{ij}$ is the sum of bridges between particle $i$ and $j$ ($j\neq i$), $n_{ij}=n^{(AC)}_{ij}$ if $\sigma_i \neq \sigma_j$ and $n_{ij}=n^{(AB)}_{ij}+n^{(AB)}_{ji}$ if $\sigma_i=\sigma_j=0$. $f^{A,B,C}_\mathrm{rep}$ is the free energy in the absence of any linkages ($n^{(XY)}_{ij}=0$, $\forall \, X,\,Y,\,i,\,j$). For thin, rigid linkers with a length $L$ much smaller than the radius of the colloids $R$, $f^{A,B,C}_\mathrm{rep}$ is the sum of repulsive, osmotic contributions written in terms of the volume excluded to the tip of linkers on particle $i$ by the presence of particle $j$, $\Omega^{\mathrm{exc}}_{ij}$ ($\Omega^{\mathrm{exc}}_{ij}=0$ for the configuration of Fig.~\ref{FigModel}). Similarly, the free energy due to inert linkers, $f_\mathrm{inert}$, is written in terms of the volume excluded to the tip of the $I$ linkers tethered to $i$ by the hard-core of particle $j$, $\Omega^{\mathrm{exc},I}_{ij}$ (see Fig.~\ref{FigModel}). In this work the length of the inert linker is taken equal to $L_I=2 \cdot L$ with colloids' radius equal to $R=5\cdot L$, and $f_\mathrm{inert}$ is approximated using pair interactions calculated for two isolated colloids.
\\ 
The forces between particles are then calculated in terms of the number of free and bound linkers and the configurational volumes available to the different types of linkers, either bound ($\Omega_{ij}$) or free ($\Omega_i$). In particular, the force acting on particle $i$ reads as \cite{A-UbertiPRL2014,PetitzonSoftMatter2016,JanaNanoscale2019}
\begin{eqnarray}
{ {\bf f}_{i} \over k_B T} &=& \sum_j\Bigg[ n_{ij} { {\bf \nabla}_i \Omega_{ij} (\{ {\bf r} \}) \over \Omega_{ij} (\{ {\bf r} \}) } - n_i { {\bf \nabla}_i \Omega^{(\mathrm{excl})}_{ij} (\{ {\bf r} \}) \over \Omega_i (\{ {\bf r} \}) } - n_j { {\bf \nabla}_i \Omega^{(\mathrm{excl})}_{ij} (\{ {\bf r} \}) \over \Omega_j (\{ {\bf r} \}) }  
\nonumber \\
&& \quad - \sigma_i N_I { {\bf \nabla}_i \Omega^{(\mathrm{excl},I)}_{ij} (\{ {\bf r} \}) \over \Omega^{(\mathrm{I})}_i (\{ {\bf r} \}) }+\sigma_j N_I { {\bf \nabla}_i \Omega^{(\mathrm{excl},I)}_{ij} (\{ {\bf r} \}) \over \Omega^{(\mathrm{I})}_j (\{ {\bf r} \}) }\Big]
\label{Eq:forces}
\end{eqnarray}
where $n_i$ is the total number of unpaired reactive linkers and loops present on particle $i$ \cite{JanaNanoscale2019}, and ${\bf \nabla}_i = \partial/\partial {\bf r}_i$. We calculate the most likely number of linkages using chemical equilibrium conditions \cite{VarillyJCP2012}
\begin{eqnarray}
{ n^{(AB)}_{jj}\over \Omega_j(\{ {\bf r} \})} &=& {1\over \rho_\ominus} { n^{(A)}_j \over \Omega_j(\{ {\bf r} \})} {n^{(B)}_j \over \Omega_j(\{ {\bf r} \})}  e^{ - {\Delta G^{(0)}_{AB} \over k_B T} }  
\nonumber \\
{n^{(AC)}_{ji}\over \Omega_{ij}(\{ {\bf r} \})} &=& {1\over \rho_\ominus}  {n^{(A)}_j\over \Omega_j(\{ {\bf r} \})} { n^{(C)}_i \over \Omega_i(\{ {\bf r} \})}  e^{ -{\Delta G^{(0)}_{AC} \over k_B T} }
\label{Eq:linkages}
\end{eqnarray} 
where $\Delta G^{(0)}_{XY}$ is the hybridization free energy of binding the reactive sequences of $X$ and $Y$ when free in solution (not tethered to any colloid), defined using the standard concentration $\rho_\ominus$, $\rho_\ominus=0.6022\,$nm$^{-3}$. 
In the following we offset $\Delta G^{(0)}_{XY}$ by $k_B T \log (\rho_\ominus L^3)$, where $L$ plays the role of the simulation unit length.
Notice that in Eq.~\ref{Eq:linkages} $n^{(X)}_i/\Omega_i$ and $n^{(AC)}_{ij}/\Omega_{ij}$ are the local densities of the tips of free and bound linkers (which are assumed to be uniform, as is the case for small values of $L/R$ and for ideal linkers \cite{A-UbertiPRL2014}). Notice also that the configurational volumes of loops and free linkers are equal ($\Omega_i=\Omega_{ii}$). We calculate $n^{(AB)}_{ij}$ and $n^{(BA)}_{ij}$ using expressions similar to Eqs.~\ref{Eq:linkages}.
\\
Importantly,  from Eqs.~\ref{Eq:linkages}, it follows that the different numbers of linkages featured by a given particle are correlated quantities. This observation is peculiar to colloids featuring mobile linkers and underlies the fact that, in these systems, particle interactions are multibody \cite{A-UbertiPRL2014,MognettiRPP2019}.

\section{Results and Discussion}

\begin{figure*}[h]
\centering
 \includegraphics[scale=0.5]{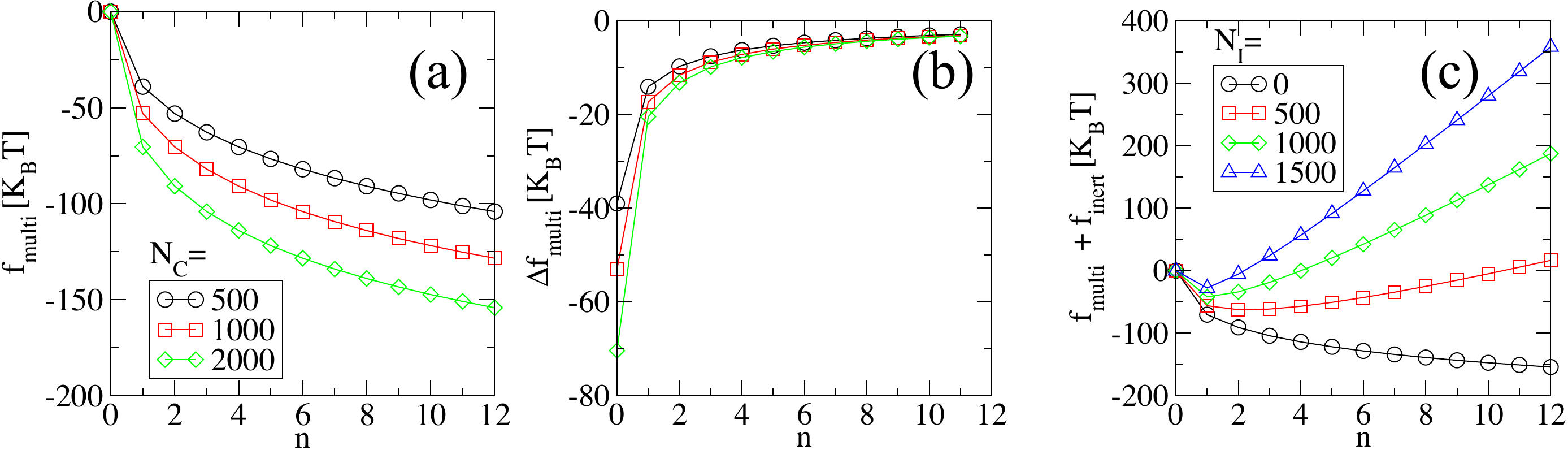} 
\caption{
(a) Multivalent free-energy of clusters made by an R particle surrounded by $n$ G particles (RG$_n$) for three different values of the number of linkers of type C $N_c$. $f_\mathrm{multi}$ is non-additive, as highlighted by the free energy gain of adding the n$^\mathrm{th}$ particle to the cluster, $\Delta f$ (b). (c) When adding inert linkers, the total free energy of the cluster becomes non-monotonous in $n$. We use $\Delta G^{(0)}_{AB}=\Delta G^{(0)}_{AC}=-9 k_B T$, $N_R=40$, and in panel (c) $N_C=2000$.
}\label{FigTheory1}
\end{figure*}

\subsection{Design rules for yielding finite-sized aggregates }

In this section, we leverage state--of--the--art multivalent theories \cite{MognettiRPP2019} to fine-tune the system’s parameters resulting in sought structures. As sketched in Fig.~\ref{FigIntro}c, we aim at fabricating clusters made of a single G particle enrobed by a controllable number of R particles. Achieving these structures requires programming interactions between particles so that:
\begin{itemize}

\item[1.] Interactions between R particles should be weak enough to avoid aggregation in single--component suspensions of colloids of type R. We use a self--protected scheme in which intra--particle loops (see Fig.~\ref{FigModel}) prevent R particles from aggregating.

\item[2.] The formation of heterodimers (RG) should be favored, but configurations made of a single R particle in direct contact with multiple G particles should be forbidden (see bottom of Fig.~\ref{FigIntro}b). The use of G particles which are not self-protected allows binding R with G, while coating G with a proper amount of inert linkers hampers the formation of RG$_n$ clusters if $n>1$ \cite{A-UbertiPRL2014}.

\item[3.] The number of particles entering a cluster should be limited and controlled by the system’s parameters. When assembling colloidal molecules, we leverage the multibody nature of the free-energy allowing us to tune the number of R particles in direct contact with a G particle. For core-shell structures, we put at play the mechanism of Ref.~\cite{JanaNanoscale2019} in which, for sufficiently strong G-R interactions, G-bound R particles can further stabilize R particles without the need for the latter to be in direct contact with the G colloid. This because G-bound R particles are not self-protected (given that they express a free B linker for each CA linkage formed, see Fig.~\ref{FigModel}) and can then attract R particles from the bulk. Importantly, the number of free B linkers expressed by R particles decreases with the number of layers limiting the growth of the cluster to a controllable size \cite{JanaNanoscale2019}.

\end{itemize}

Notice that the two key properties limiting the growth of the aggregates are the possibility of barring the formation of $RG_2$ trimers (design rule 2, above) and the control over the number of  R particles attaching a cluster (design rule 3, above).  Below we report free energy calculations of typical colloidal configurations and clarify how to translate the previous design rules into precise sets of the system parameters. 

1. {\bf Single component suspensions of R colloids.} We decorate R particles with two types of complementary linkers (A and B, see Fig.~\ref{FigModel}). The phase behavior of the system is controlled by the competition between inter-particle bridges and intra-particle loops. At low values of the hybridization free energy of the sticky ends $\Delta G^{(0)}_{AB}$ (see Eq.~\ref{Eq:linkages}), all linkers are paired forming either a loop or a bridge. In particular, particles in the assembled and gas phase present the same number of reacted linkers. It follows that, at low temperature $T$ (corresponding to low values of $\Delta G^{(0)}_{AB}$), the phase boundary is not a function of $T$ (in particular, there are systems that do not phase separate even at very low $T$ \cite{Sciortino2019NuovoCimento}). Instead, the thermodynamically stable phase is determined by entropic contributions, namely, the combinatorial term counting the number of ways of reacting the linkers, the configurational costs of forming bridges and loops (see Eq.~\ref{Eq:linkages}), and the chemical potential of the colloids \cite{VarillyJCP2012,RogersNatRevMat2016,Sciortino2019NuovoCimento}. For a given radius $R$ and length of the linkers $L$, aggregation is induced either by increasing the number of linkers per particle $N_R$ or by increasing the chemical potential of the colloids \cite{BachmannSoftMatter2016,JanaNanoscale2019}. 
\\ 
Ref.~\cite{JanaNanoscale2019} studied in detail the phase behavior of suspensions of R particles and reported results in agreement with experiments using large unilamellar vesicles. Similar phase behaviors have been found in systems featuring linkers with fixed tethering points \cite{A-UbertiNatMat2015,RogersScience2015,RogersNatRevMat2016}. In the present study, we choose values of $N_R$ and chemical potential (or density of the particles) resulting in stable gas phases.

2. {\bf Free energy of RG$_n$ clusters.} In this paragraph, we assess the strategy to hamper configurations made of two G colloids cross-linked by an R particle (see design rule 2 above and Fig.~\ref{FigIntro}b) using the free energy detailed in the method section. In Fig.~\ref{FigTheory1}, we consider the free energy of configurations made of an R particle surrounded by $n$ G particles ($n=$1, …, 12). We place R on a site of an FCC lattice and add G particles to the neighboring sites of R, sequentially filling the three planes orthogonal to the [111] direction (starting from the plane containing R). We fix all the distances between neighboring particles to $d=11\cdot L$, comparable with the distance that minimizes the multivalent free energy ($f_\mathrm{multi}$, see Eq.~\ref{Eq:fmulti}) calculated for two particles. Fig.~\ref{FigTheory1} reports $f_\mathrm{multi}$ of the system for three different numbers of C linkers on particle G. (Importantly, in Fig.~\ref{FigTheory1} and the following, we subtract to $f_\mathrm{multi}$ the reference free-energy consisting of isolated particles which, in particular, could feature loops.) We find that $f_\mathrm{multi}$ is non-additive in $n$. This trend is better highlighted in Fig.~\ref{FigTheory1}b where we report the change in free-energy of adding a G particle to a cluster containing $n-1$ G colloids, $\Delta f_\mathrm{multi}(n)=f_\mathrm{multi}(n)- f_\mathrm{multi}(n-1)$. $\Delta f_\mathrm{multi}$ steadily increases with $n$. This result is because G particles compete to bind the A strands present on the R particle, and such competition is more severe for high values of $n$. The results of Fig.~\ref{FigTheory1} enable remote control over the number of G particles bound to an R one.  Valency control is achieved by adding inert linkers to the G particles to destabilize each G-R contact \cite{A-UbertiPRL2014}. In particular, if the inert linker contribution per pair of colloids becomes comparable or bigger than $\Delta f_\mathrm{multi}(2)$, then RG$_2$ (and similarly RG$_n$, with $n>2$) becomes unstable. The previous estimation neglects the entopic losses of the centers of mass of the colloids forming a molecule. Such contributions, which are negligible if $\Delta f_\mathrm{multi}(n)$ is steep (as compared to $k_B T n$), are sampled using numerical simulations (see the next section). In Fig.~\ref{FigTheory1}c, we consider the total free energy of the system at different numbers of inert linkers $N_I$. Values of $N_I$ bigger than $N_I=1500$ are sufficient to bar the formation of RG$_2$ molecules. Importantly, the number of inert linkers necessary to destabilize RG$_2$ depends on the system’s parameters, namely, the number of linkers and the hybridization free energies. Notice that, as compared to the design of Ref.~\cite{A-UbertiPRL2014}, inert linkers are only on one type of particle. In such a way we do not limit the number of R particles bound to G opening the avenue to the formation of colloidal and core-shell structures.

\begin{figure}[h]
\centering
 \includegraphics[scale=0.4]{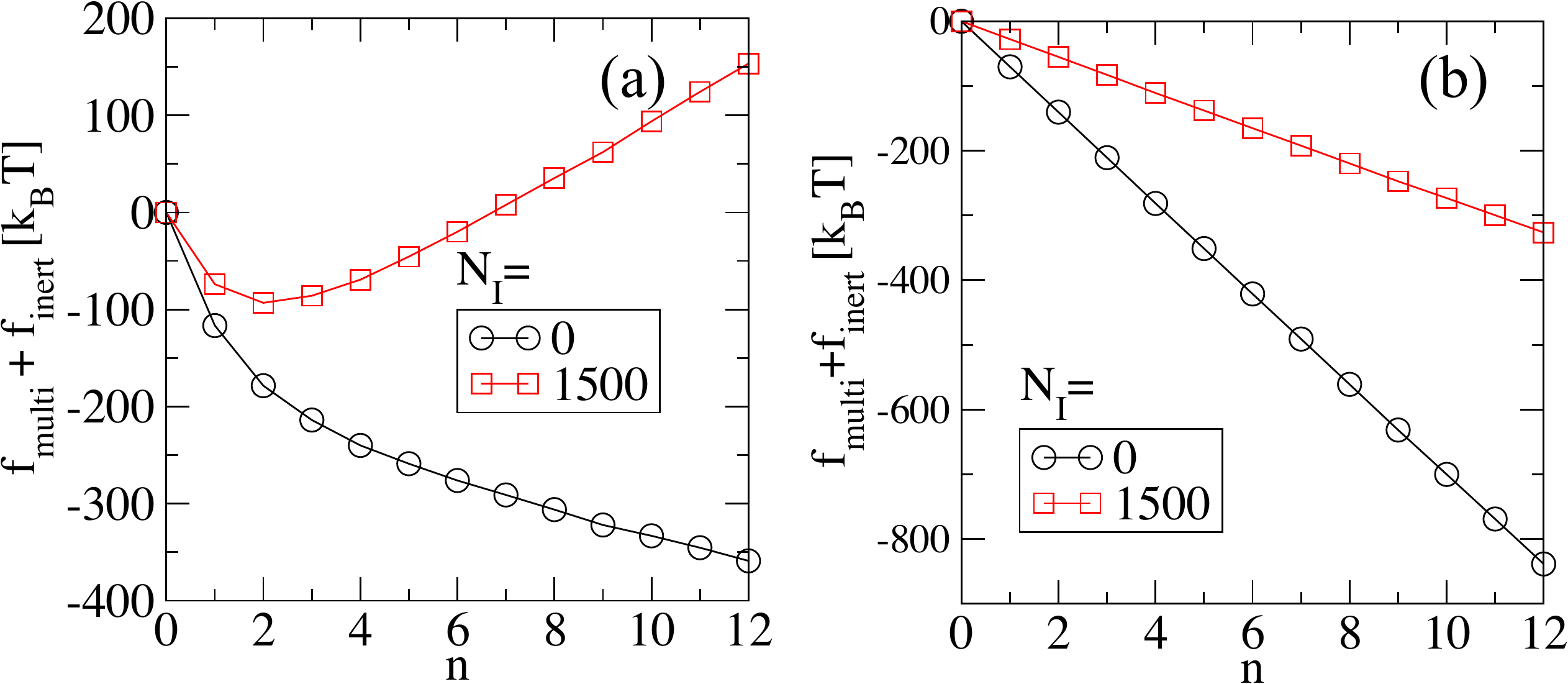} 
\caption{Free energy $f$ of clusters made by a G particle surrounded by $n$ R particles. (a) When using a comparable number of C and A/B linkers ($N_C=N_R=40$), the free energy is non-monotonous leading to self-assembly of colloidal molecules with a controllable valency. (b) When increasing the number of C linkers ($N_C=2000$ and $N_R=40$) the free energy becomes linear and the system self-assemble core-shell structures. In (a) we use $\Delta G^{(0)}_{AB}= -9 k_B T$ and $\Delta G^{(0)}_{AC}=-15 k_B T$, while in (b) $\Delta G^{(0)}_{AB}= \Delta G^{(0)}_{AC} = -9 k_B T$.
}\label{FigTheory2}
\end{figure}

3. {\bf Free energy of GR$_n$ clusters.} The design rules 1 and 2 allow assembling colloidal clusters made up of at most a single G particle through fine-tuning the number of linkers on R particles (rule 1) and the number of inert linkers on G particles (rule 2). We now show how, by changing the number of C linkers on G particles, it is possible to control the number of R particles entering the cluster, therefore allowing to self assemble colloidal molecules, GR$_z$, with a controlled valency $z$. In Fig.~\ref{FigTheory2} we study the free energy $f$ of a cluster made of a G particle and $n$ R particles. For a small number of C linkers on the G particle (Fig.~\ref{FigTheory2}a), $f_\mathrm{multi}$ (corresponding to the curve with $N_I$=0) shows a non-linear behavior (as discussed in Fig.~\ref{FigTheory1}). The linear trend of $f_\mathrm{multi}$ (and $f=f_\mathrm{multi}+f_\mathrm{inert}$) is recovered at a high number of C linkers (see Fig.~\ref{FigTheory2}b), as in such condition the depletion of C linkers following the binding of R particles becomes negligible. In the presence of inert linkers and small numbers of C linkers (Fig.~\ref{FigTheory2}a),  $f$ develops a minimum allowing to tuning the valency of the colloidal molecule. 
At a high number of C linkers (Fig.~\ref{FigTheory2}b), the G colloids are enrobed by a shell of closed packed R colloids. In this case we can leverage the effect described in Ref.~\cite{JanaNanoscale2019}, and try to assemble a second shell of R particles by further refining the number of linkers on the R particle or the density of R colloids. The possibility of assembling core-shell aggregates with multiple shells of R colloids is demonstrated using simulations in the next section. \\
We should warn that in this section we have neglected residual interactions between R particles belonging to different clusters. Such interactions may (reversibly) cross-link different clusters (both in the case of molecules and core-shell structures). As proven in the next sections, cross-linking between different clusters is suppressed by lowering the packing fraction of the system. 

\begin{figure*}[h!]
\centering
 \includegraphics[scale=0.5]{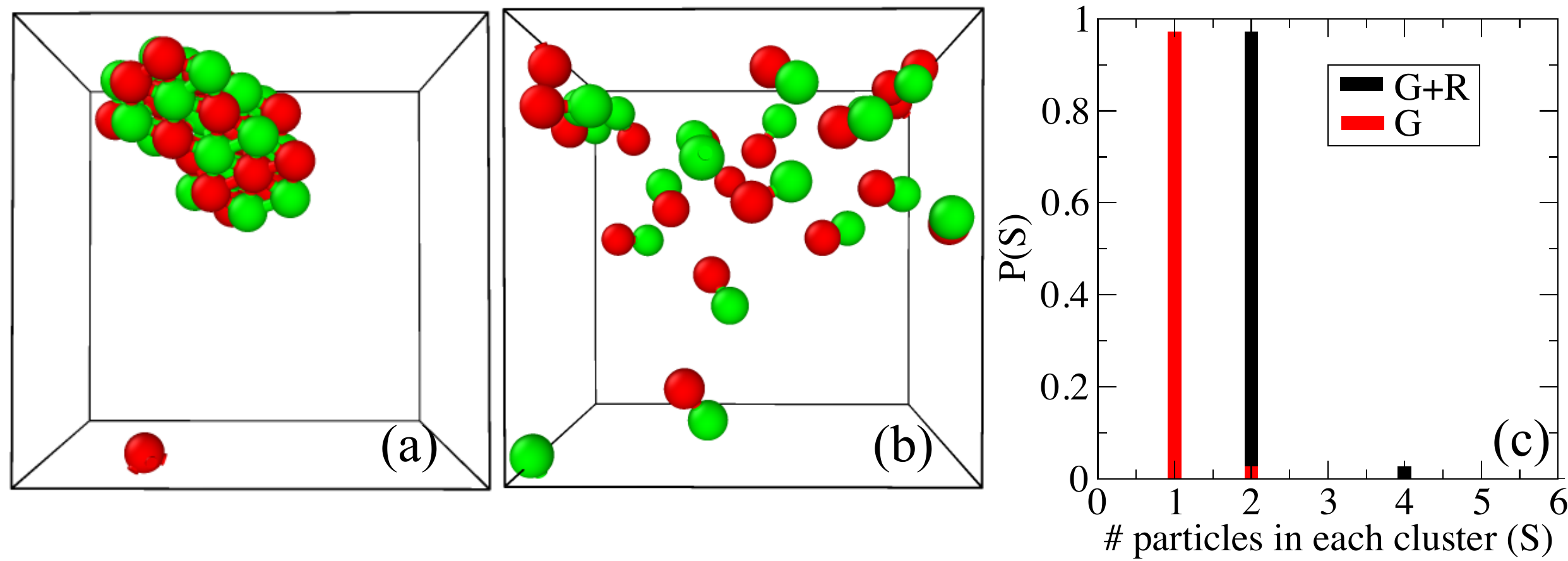} 
\caption{ Simulation snapshots in steady conditions of the system of Fig.~\ref{FigTheory2}a  with $N_I=0$ (a) and $N_I=1500$ (b). We use 20 G and 20 R colloids placed in a cubic box with side length equal to 100$\cdot L$. (c) Probability distributions of the total number of colloids (G+R) and the number of G colloids (G) per cluster obtained in steady conditions.
}\label{FigSim1}
\end{figure*}

\subsection{Simulation results}

To verify the theoretical predictions, we perform computer simulations in which we update the number of linkages $\{n\}$ and colloids’ positions $\{\bf{r}\}$ in a concerted way. At each step of the simulation scheme, we first calculate the most likely number of inter and intra-particle linkages using Eq.~\ref{Eq:linkages}, keeping $\{\bf{r}\}$ fixed. Then, we calculate the forces acting on each particle ${\bf{f}}_i$ using Eq.~\ref{Eq:forces}. Using ${\bf{f}}_i$, we update the
positions of the particles using a Brownian dynamics scheme
\begin{equation}
{\bf{r}}_i(t+\Delta t)={\bf{r}}_i(t)+{\bf{f}}_i\frac{D}{k_BT}\Delta t+\sqrt{2D\Delta
t}\cdot{{\bf W}_i}(0,1)
\end{equation}
where $D$ is the particle diffusion constant in the diluted limit. $\Delta t$ is the integration time step and ${\bf W}_i$ is a normal distributed vector with covariance matrix equal to the identity matrix. All simulations are performed at a constant total number of particles, $N_p$, and temperature $T$. The unit of length and time are $L$ and $L^2/D$.

In Fig.~\ref{FigSim1}, we demonstrate the importance of using inert linkers to yield finite-sized aggregates. To this end, we consider the model of Fig.~\ref{FigTheory2}a and compare the case in which G particles carry no inert linkers (Fig.~\ref{FigSim1}a) with the one in which $N_I=1500$ (Fig.~\ref{FigSim1}b). The snapshots of Fig.~\ref{FigSim1} are taken in steady conditions. Fig.~\ref{FigSim1} shows how, without inert linkers, all the colloids precipitate into a single aggregate as found in systems undergoing a first-order phase transition. Instead, in the presence of inert linkers, as predicted by the design rule 2 (see the previous section), finite-sized aggregates appear. We confirm this claim in Fig.~\ref{FigSim1}c where we perform a cluster analysis and report the total number of particles {\em per} cluster (R+G) and the number of G particles in each cluster (G). Almost all clusters contain at most a single G particle as predicted by theory.

\begin{figure}[h]
\centering
 \includegraphics[scale=0.5]{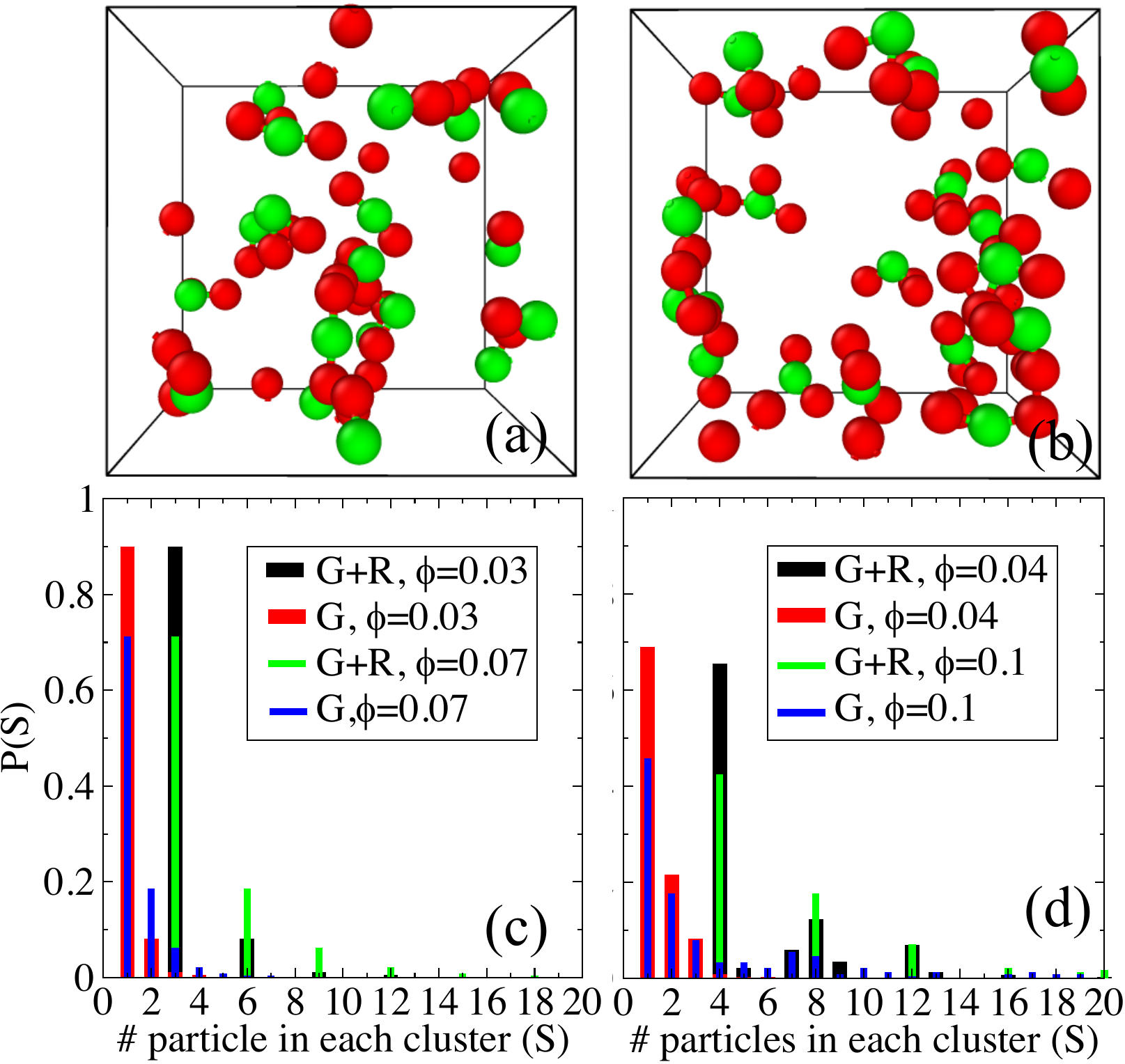} 
\caption{ Self-assembly of trimers (a, c) and tetramers (b, e) starting from suspensions with 20 G particles and, respectively, 40 and 60 R particles. The snapshots have been taken in steady conditions at packing fraction equal to $\phi=0.03$ (a) and $\phi=0.04$ (b). Panels (c) and (d) report the cluster analysis similar to what reported in Fig.~\ref{FigSim1}. The model parameters are as in Fig.~\ref{FigTheory2}a with $N_I=1500$.
}\label{FigSim2}
\end{figure}

Fig.~\ref{FigSim1}c shows how the system features a tiny fraction of clusters made by two G particles and two R particles (resulting in 4 total particles). These clusters resemble colloidal heteropolymers (with sequence GRRG) in which two bound R particles cross-link two G particles and result from residual interactions between G-bound R particles (discussed in the design rule 3, see previous section). These higher-order structures are transient states, as demonstrated in Fig.~\ref{FigSim2}. In Fig. 6 we consider the model of Fig.~\ref{FigTheory2} for two different stoichiometric ratios: [R]:[G]=1:2 (Fig.~\ref{FigSim2}a and \ref{FigSim2}c) and [G]:[R]=1:3 (Fig.~\ref{FigSim2}b and \ref{FigSim2}d) at different packing fractions (see legends of panel c and d). Considering the [R]:[G]=1:2 case, Fig.~\ref{FigSim2}c shows that, as expected, the system yields mostly trimers made of one G particle and two R particles (G+R=3).  Fig.~\ref{FigSim2}c  also shows how the number of spurious structures made by two G particles and 4 R particles decreases at low packing fraction. This result is because, at low packing fraction, the encounter rate between molecules is lower and proves the fact that higher-order structures are not stable. When considering tetramers (Fig.~\ref{FigSim2}d), we find that at high packing fractions ($\phi=0.1$) the distribution of G particles {\em per} cluster broaden with transient clusters counting up to 12 particles. At lower packing fraction ($\phi=0.04$) the systems predominantly feature tetramers and pairs of tetramers reversibly attached. Taken together, the results of Fig.~\ref{FigSim2}c and \ref{FigSim2}d show how one may need to fine-tune the packing fraction of the system to optimize the production of desired finite-sized clusters. Alternatively, one could weaken the interactions between G-bound R colloids by decreasing the number of A and B linkers tethered to R particles ($N_R$) \cite{JanaNanoscale2019}. 

\begin{figure}[h]
\centering
 \includegraphics[scale=0.4]{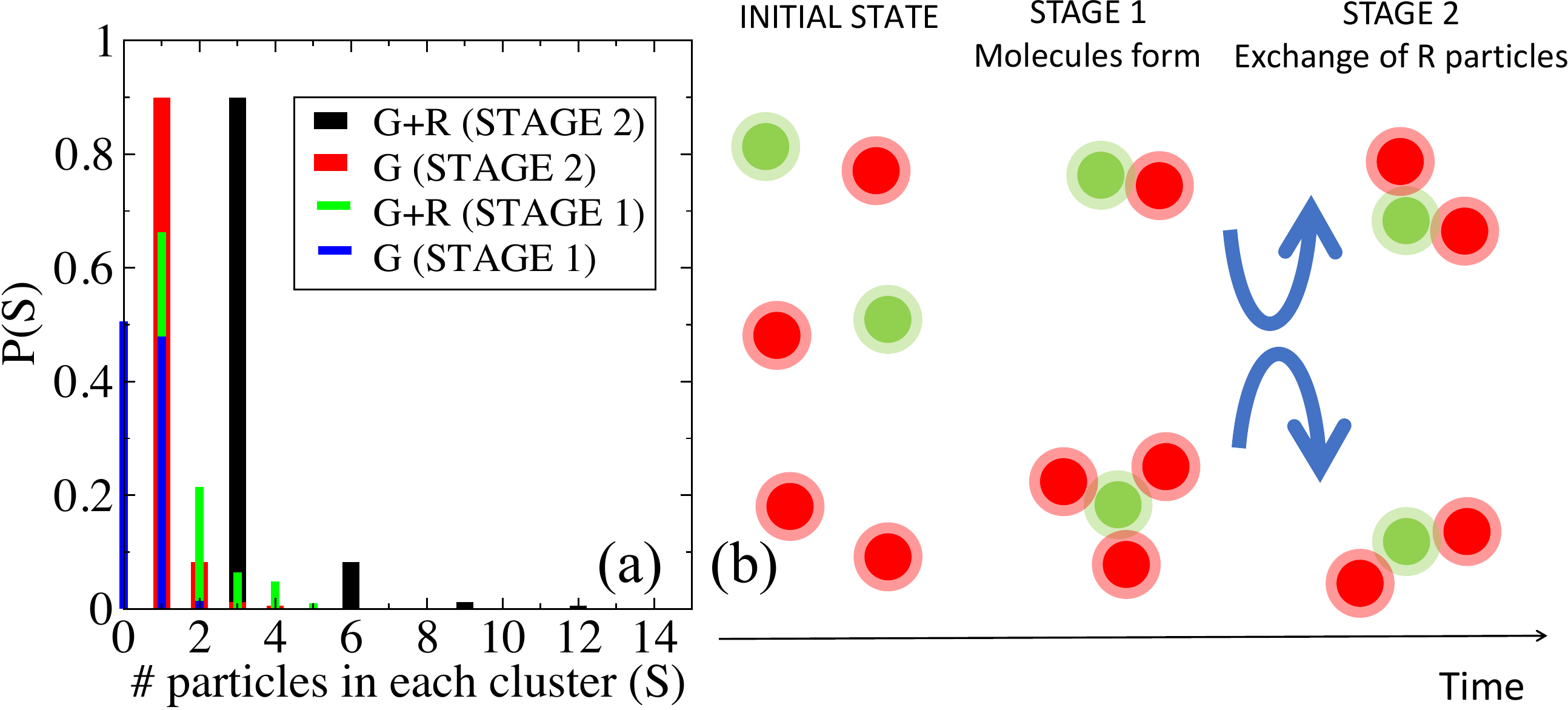} 
\caption{ (a) Cluster analysis of the system of Fig. 4a with [R]:[G]=2:1 in the early stages of aggregation (STAGE1) and steady conditions (STAGE2). (b) In STAGE1, the system form molecules with a broad distribution of the number of R particles. Before reaching steady conditions (STAGE2), R particles are redistributed between colloidal molecules. We use 20 G and 40 R particles with packing fraction equal to $\phi=0.031$. The number of inert linkers is $N_I=1500$.
}\label{FigSim3}
\end{figure}

Figs.~\ref{FigSim1} and \ref{FigSim2} show how R particles are evenly distributed among G colloids resulting in colloidal molecules with a well-defined valency, $z$, corresponding to the stoichiometric ratio of the two types of particles, $z=$[R]:[G].  This result may look puzzling since density fluctuations of R particles should likely result in different G particles capturing different numbers of R particles. In fact, the system does not feature a broad distribution of valencies since the free energy of GR$_n$ clusters is non-linear in $n$ (see Fig~\ref{FigTheory2}a). In particular, the ground state of the system corresponds to configurations with molecules featuring the same number of R particles. 
\\
To better understand the self-assembly pathway leading to colloidal molecules, in Fig.~\ref{FigSim3}a we report the cluster analysis of a system forming trimers ([G]:[R]=1:2) in the early stages of aggregation (STAGE1) as compared to the cluster analysis in steady conditions (STAGE2).   In STAGE1, as in steady conditions, two G particles rarely belong to the same cluster because of design rule 2 (see previous section). However, at this stage, the number of particles {\em per} cluster is broadly distributed with the system featuring dimers, trimers, and tetramers (see STAGE1 in Fig.~\ref{FigSim3}b). While reaching steady conditions (STAGE2), R particles are shuffled between different molecules. Redistribution of R particles happens during molecular collisions, as depicted in Fig.~\ref{FigSim3}b. Using the results of the previous section (see Fig.~\ref{FigTheory2}), it is possible to calculate the free energy gain, $\delta f$, of redistributing R particles between G particles. For the exchange of one R particle from a tetramer to a dimer (see Fig.~\ref{FigSim3}b) we have (using the notation of the previous section):
\begin{eqnarray}
\delta f &=& 2 f(2) - f(3) - f(1) 
\nonumber \\
&=&  (f_\mathrm{multi}(2)-f_\mathrm{multi}(1)) - (f_\mathrm{multi}(3)-f_\mathrm{multi}(2))<0 
\nonumber
\end{eqnarray}
where the latter inequality follows from the  fact that $(f_\mathrm{multi}(n)-f_\mathrm{multi}(n-1))$  increases with $n$ (similarly to what happens in  Fig.~\ref{FigTheory1}a and \ref{FigTheory1}b for RG$_n$ clusters).

\begin{figure*}[h]
\centering
 \includegraphics[scale=0.5]{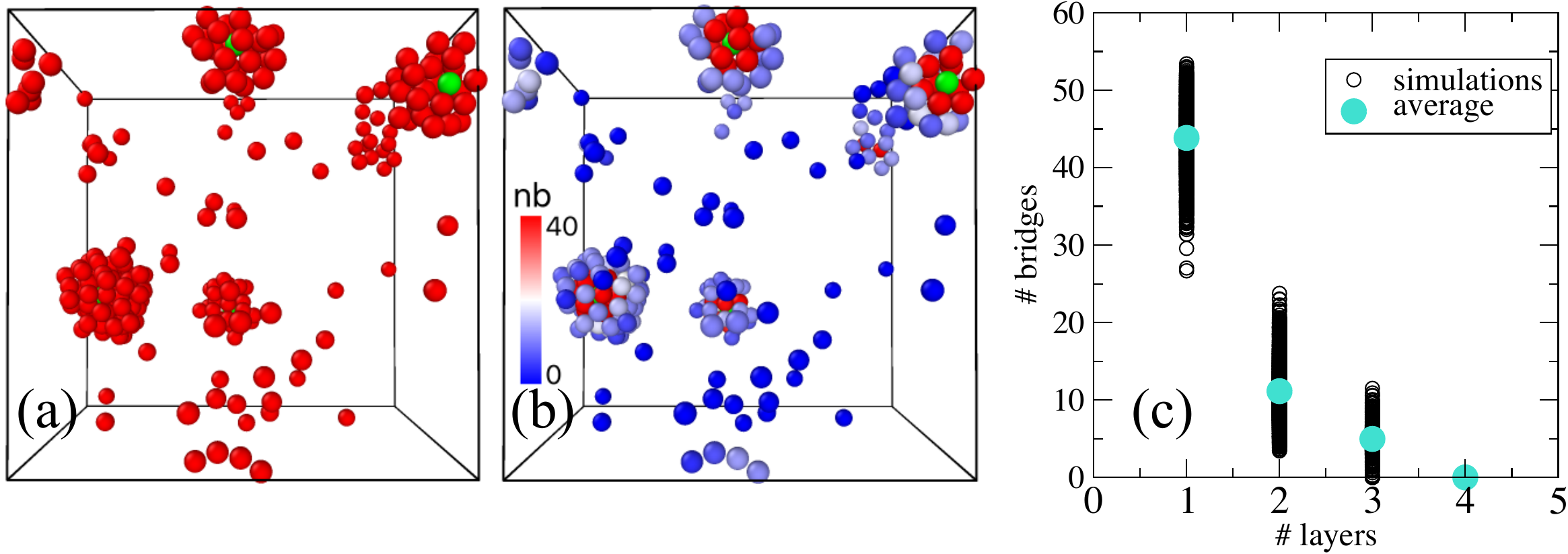} 
\caption{ (a) Core-shell structures observed in steady conditions for the system of Fig. 4b with 5 R particles and 250 R particles in a cubic box with side equal to 200 L. The number of inert linkers is equal to $N_I=1500$. (b) For the configuration of panel (a), we report the number of inter-particle bridges featured by R particles. (c) Simulation data and averages of the number of bridges of R particles as a function of the distance of the layer on which the R particle lays.
}\label{FigSim4}
\end{figure*}

Finally, we demonstrate the ability of our system to self-assemble core-shell structures. We use the model of Fig.~\ref{FigTheory2}b in which the free energy of RG$_n$ clusters is linear in $n$, resulting in G colloids enrobed by a compact shell of R particles.  Intriguingly, our design allows growing a second shell of R particles as demonstrated in the snapshot of Fig.~\ref{FigSim4}a.  In particular, particles on the first shell can stabilize  R particles from the bulk without the necessity for the latter to be in direct contact with the G particle at the center of the cluster. This observation is in apparent contrast with the fact that R particles do not aggregate in bulk (design rule 1). In fact, R particles in bulk are protected by intra-particle loops between A and B linkers, while G-bound particles feature free B linkers given that A linkers are employed to form inter-particle bridges between R and G (see Fig.~\ref{FigModel}). Fig.~\ref{FigSim4}b and \ref{FigSim4}c show how the number of bridges featured by R particles decreases as a function of the distance from the G particles. At a certain point, the number of bridges becomes too small and the growth of the aggregate comes to a halt. This mechanism has been introduced by Ref.~\cite{JanaNanoscale2019} that also explained how to control the number of layers by controlling the number of  A/B linkers ($N_R$) and the chemical potential of the R particles. As observed for colloidal molecules, core-shell structures may cross-link each other through residual interactions between R particles. Such conditions could be avoided through the use of highly asymmetric stoichiometric ratios (as in Fig.~\ref{FigSim4}) or by further refining the number or A/B linkers. A detailed investigation of the thermodynamic conditions avoiding cross-linking between core-shell structures deserves future investigations.

\section{Conclusions}

The last 20 years have witnessed much progress in the field of colloidal self-assembly directed by DNA oligomers. DNA allows programming a myriad of orthogonal interactions \cite{JonesScience2015,RogersNatRevMat2016} and provides direct control over the morphological properties of both disordered \cite{VarratoPNAS2012,DiMicheleSoftMatter2014} and ordered aggregates (crystal type, size of the unit cell, … ) \cite{JonesScience2015,Ducrot_NatMat_2017,Wang_NatComm_2017,Liu_Science_2016,AuyeungNature2014,NykypanchukNatMat2008}. Importantly, recent work has also clarified the kinetics of colloids interacting through reacting linkers resulting in design principles for avoiding kinetic bottlenecks hampering the relaxation of the system towards the sought target \cite{WangNatComm2015,LeeSoftMatter2018,JanaPRE2019}.
\\
So far, most of the work on DNA-mediated interactions has focused on the self-assembly of bulk, extended structures. When aiming at self-assembling finite-sized structures from isotropic unit components, one faces the necessity of controlling the number and type of the particles entering each aggregate. Spatial sorting of colloids at the microscopic scale is not possible when starting from diluted suspensions, which usually feature large density and stoichiometric fluctuations. 
\\
In this paper, we have addressed this challenge by designing a new multivalent system yielding thermodynamically stable finite-sized aggregates. We have studied a system made by two types of colloids (G and R). We have programmed the interactions between particles so that, in the early stages of the colloidal aggregation, G can bind R particles while R particles cannot cross-link two (or multiple) G particles.  The latter restriction bends the self-assembly pathway towards the formation of finite-sized clusters seeded by single G particles surrounded by R colloids. Moreover, our design allows controlling the number of R particles resulting in colloidal molecules or core-shell structures. Importantly, in the former case, we yield molecules with a well-defined valency. Such an extraordinary level of controllability of the morphological properties of the assembly relies on the underlying multibody interactions featured by interactions mediated by mobile linkers in combination with non-specific, repulsive interactions (in this paper modeled using inert linkers). It should be noticed that, in the past, pair-potentials featuring short-ranged attractive wells and repulsive shoulders have been used to self-assemble open structures \cite{GroenewoldJPCB2001,GroenewoldJPCM2004,Dinsmore2011}. As compared to the latter work, in the present model repulsive interactions are short-ranged. Instead, as for the design of Ref.~\cite{A-UbertiJCP2013}, valency control arises from the fact that the attractive part of the multivalent free energy {\em per} pair of bound particles becomes weaker at high valency while the repulsive term stays constant.  
\\
We have provided design rules linking the system’s parameters to the targeted structure that have been validated using numerical simulations. Importantly, we have assumed that the timescales at which linkages form and break are much smaller than the typical timescales at which particles diffuse. If, on the one side, this assumption is not always justified \cite{PetitzonSoftMatter2016,LanfrancoLangmuir2019}, DNA sequence designs based on the toehold-exchange mechanism \cite{ZhangJACS2009} allow speeding up the reaction kinetics by orders of magnitudes \cite{ParoliniACSNano2016}. In particular, the sequences employed in Ref.~\cite{ParoliniACSNano2016} could be readily used in the present design to speed up the conversion between inter-particle and intra-particle linkages. 
\\
Beyond the field of programmable self-assembly, this paper expands the ensemble of phase behaviors featured by multivalent systems and may shed new light on critical biological systems. For instance, the nature of the driving forces underlying the formation of membraneless aggregates made by proteins in cells is not well understood and is currently debated \cite{BerryRPP2018}. Our work shows how microphase separation may result from thermodynamic principles other than being controlled by kinetic factors \cite{BrachaCell2018,RanganathanArxiv2020}.

\section*{Acknowledgements}
This work was supported by the Fonds de la Recherche
Scientifique de Belgique (F.R.S.-FNRS) under grant n$^\circ$ MIS
F.4534.17 and by an ARC (ULB) grant of the {\em F\'ed\'eration Wallonie-Bruxelles}. 
Computational resources have been provided by the {\em Consortium des \'Equipements de Calcul Intensif} (C\'ECI), funded by F.R.S.-FNRS under Grant n$^\circ$ 2.5020.11 and by the Walloon Region.

\end{document}